\documentclass[prd,twocolumn,showpacs,aps]{revtex4-1}

\usepackage{amsmath, amsfonts, amssymb, mathrsfs}
\usepackage{txfonts}
\usepackage{graphicx}
\usepackage{epsfig}
\usepackage{enumerate}
\usepackage{caption}
\usepackage{subcaption}

\usepackage{slashed}
\usepackage{ulem}

\usepackage{ifpdf}
\ifpdf
\usepackage{epstopdf}
\fi

\usepackage[unicode]{hyperref}
\hypersetup{
	pdftitle={Gravitational waves of nonextremal Kerr black holes from conformal symmetry},
	pdfauthor={Nian, Tian},
	pdfsubject={Gravitational waves of nonextremal Kerr black holes from conformal symmetry},
	colorlinks=true,
	linkcolor=blue,
	citecolor=blue,
	filecolor=black,
	urlcolor=blue
}

\def\url#1{}

\usepackage{bm}
\usepackage{color}

\begin{document}

\title{Gravitational waves of nonextremal Kerr black holes from conformal symmetry}
\author{Jun Nian$^1$}
\email{nianjun@ucas.ac.cn}
\author{Weijie Tian$^1$}
\email{tianweijie19@mails.ucas.ac.cn}
\affiliation{$^1$International Centre for Theoretical Physics Asia-Pacific, University of Chinese Academy of Sciences, 100190 Beijing, China}

\begin{abstract}
In the low-energy limit,  the near region of a generic Kerr black hole has been conjectured to be holographically dual to a two-dimensional conformal field theory.  In this paper,  we consider a test object orbiting in the near region of a nonextremal Kerr black hole.  We first couple it to a massless scalar field.  The resulting scalar radiation at the horizon is computed from the perspectives of gravity and dual conformal field theory.  Considering the influence of nonzero temperature,  the agreement of both computations is found.  Then,  we generalize the analysis to the gravitational radiation and find agreement again.  The agreement supports the conjectured holography and provides a potential theoretical tool for gravitational wave computations.
\end{abstract}

\maketitle

\section{Introduction}

Black holes are so fascinating that a lot of effort has been put into understanding them.  One of the most significant breakthroughs is the discovery of the Bekenstein-Hawking entropy,  which was inspired by the analogy between the laws of black hole dynamics and the laws of thermodynamics \cite{Christodoulou:1970wf,  Christodoulou:1971pcn,  Penrose:1971uk,  Hawking:1971tu,  WOS:A1972N081300008,  Bekenstein:1973ur,  Bardeen:1973gs,  Bekenstein:1974ax,  Hawking:1974rv,  Hawking:1975vcx}.  A direct way to explain the microscopic origin of the black hole entropy is the quantum theory of gravity,  such as string theory,  and it succeeded for some classes of black holes \cite{Strominger:1996sh,  Callan:1996dv}.  However,  a universal statistical mechanical interpretation of the Bekenstein-Hawking entropy is still missing.  A promising way for this is the holographic principle \cite{tHooft:1993dmi,  Susskind:1994vu,  Maldacena:1997re},  of which the validity is independent of the details of quantum gravity.  For example,  for black holes whose near-horizon regions are locally three-dimensional anti-de-Sitter ($\mathrm{AdS_3}$) space,  the derivation of the entropy through holographic duality applies to any consistent,  unitary quantum theory of gravity \cite{Strominger:1997eq}.

For black holes existing in the universe,  such as Kerr black holes,  things become a little different.  The simplest case might be the extreme Kerr black hole whose mass $M$ and angular momentum $J$ satisfy $J=GM^2$.  For an extreme Kerr black hole,  there is a near-horizon region,  known as the near-horizon extreme Kerr (NHEK) geometry \cite{Bardeen:1999px},  in which the asymptotic symmetry generators form the Virasoro algebra under some consistent boundary conditions \cite{Guica:2008mu}.  This motivated the Kerr/CFT correspondence \cite{Guica:2008mu},  which posits that the NHEK geometry is dual to a two-dimensional conformal field theory (CFT).  The conjecture was supported by the facts that the CFT microstate degeneracy inferred from the Cardy formula reproduces the black hole entropy \cite{Guica:2008mu} and the CFT correlators agree precisely with the scattering amplitudes of the Kerr black hole \cite{Bredberg:2009pv}.  The Kerr/CFT correspondence can be exploited to compute gravitational waves.  For an extreme or near-extremal Kerr black hole,  whether the gravitational radiation is emitted by a star in a circular orbit near the horizon or by a star plunging into the black hole,  agreements were found between the gravity and the CFT analyses \cite{Porfyriadis:2014fja,  Hadar:2014dpa,  Hadar:2015xpa}.  In fact,  through conformal symmetry,  all corotating equatorial orbits near the horizon are related,  which provides convenience for analytical computations of gravitational waves from plunges into supermassive near-extremal Kerr black holes \cite{Compere:2017hsi}.

However,  for nonextremal Kerr black holes,  the NHEK geometry disappears,  and the near-horizon conformal symmetry is not manifest.  Fortunately,  the solution space of the wave equation still has a conformal symmetry called the hidden conformal symmetry \cite{Castro:2010fd}.  In the low-energy limit $\omega M \ll 1$,  where $\omega$ is the frequency of the propagating field,  the spacetime can be divided into a near region ($r \ll 1/\omega$) and a far region ($r \gg M$),  which have an overlap.  The hidden conformal symmetry acts on the near-region wave equation at low frequencies.  Note that the near region defined above is not the near-horizon region in the definition of the NHEK geometry.  With the hidden conformal symmetry,  the Bekenstein-Hawking entropy and the scattering amplitudes of the nonextremal Kerr black hole can be reproduced \cite{Castro:2010fd},  which corroborates the (generalized)  Kerr/CFT conjecture that the near region of a generic nonextremal Kerr black hole is dual to a two-dimensional CFT.  Complementarily, the hidden conformal symmetry of a nonextremal Kerr black hole can be explicitly realized using the covariant phase space formalism \cite{Haco:2018ske}.

In this paper,  we attempt to apply the hidden conformal symmetry to gravitational wave computations of a nonextremal Kerr black hole.  We consider a star orbiting in the near region of a Kerr black hole,  whose mass is far greater than that of the star.  In the low-energy and near-region limits,  the wave equation can be solved by hypergeometric functions,  and the radiation flux at the horizon is computed.  In the dual CFT,  the orbiting star leads to a deformation of the CFT and transitions out of the thermal state.  We use Fermi's golden rule to compute the transition rate,  which is dual to the radiation flux at the horizon.  Through explicit computations,  we find agreement between results from the gravity and the CFT analyses,  which provides new evidence for the dual conformal symmetry.

This work also paves the way for the analytical study of gravitational waves through the hidden conformal symmetry.  Since LIGO first detected a gravitational wave signal from a binary black hole merger \cite{LIGOScientific:2016aoc},  gravitational wave observations have significantly developed.  Astronomical observations show many Kerr black holes spinning rapidly or slowly existing in our universe \cite{Reynolds:2020jwt}.  If there is a star orbiting near the horizon of a supermassive Kerr black hole ($\gtrsim 10^5$ solar masses),  the emitted gravitational waves ($\lesssim$ mHz),  which are the primary goal of space-based gravitational wave detections,  satisfy the low-energy limit.  The results of this paper suggest that the hidden conformal symmetry might benefit such gravitational wave computations.

This paper is organized as follows.  Section \ref{sec:review} briefly reviews the Kerr black hole metric and the hidden conformal symmetry.  In Sec.~\ref{sec:scalar},  we couple a massless scalar field to the orbiting star as a warmup.  The scalar radiation is computed from the gravity and the dual CFT perspectives,  and the results are compared.  We also reattach the far region and compute the scalar radiation at the future null infinity.  In Sec.~\ref{sec:gravi},  we adopt the Newman-Penrose formalism and generalize the results of the scalar field to the gravitational field case.  A summary and some discussions are included in Sec.~\ref{sec:discu}. Throughout this paper, we use units with $G=c=\hbar=1$.

\section{Review of Kerr black hole and hidden conformal symmetry}\label{sec:review}

In this section,  we briefly review the Kerr black hole and its dual CFT.  For a four-dimensional asymptotically flat Kerr black hole,  the metric in Boyer-Lindquist coordinates is
\begin{equation}
\begin{aligned}
ds^2 =& -\frac{\Delta}{\Sigma^2}(dt-a\sin^2\theta d\phi)^2 + \frac{\sin^2\theta}{\Sigma^2}\left[(r^2+a^2)d\phi-adt\right]^2\\
&+ \frac{\Sigma^2}{\Delta}dr^2 + \Sigma^2d\theta^2,
\end{aligned}
\end{equation}
with
\begin{equation*}
\Delta = r^2 - 2Mr + a^2,\ \Sigma^2 = r^2 + a^2\cos^2\theta,
\end{equation*}
where $M$ is the mass of the black hole,  and $J=aM$ is the angular momentum.  The inner and the outer horizons of the Kerr black hole are located at
\begin{equation}
r_\pm = M \pm \sqrt{M^2-a^2}.
\end{equation}

The Klein-Gordon equation for a massless scalar field $\Psi$ is
\begin{equation}\label{eq:KG}
\frac{1}{\sqrt{-g}}\partial_\mu(\sqrt{-g}g^{\mu\nu}\partial_\nu\Psi) = 0.
\end{equation}
In the Kerr metric,  expanding $\Psi$ in modes
\begin{equation}
\Psi(t,r,\theta,\phi) = e^{-i\omega t+im\phi}R(r)S(\theta),
\end{equation}
Eq.~\eqref{eq:KG} can be separated into
\begin{equation}\label{eq:KGtheta}
\left[\frac{1}{\sin\theta}\partial_\theta(\sin\theta\partial_\theta) - \frac{m^2}{\sin^2\theta} + a^2\omega^2\cos^2\theta\right] S(\theta) = -K_l\, S(\theta),
\end{equation}
and
\begin{equation}\label{eq:KGr}
\begin{aligned}
&\Biggl[\partial_r(\Delta\partial_r) + \frac{(2Mr_+\omega-am)^2}{(r-r_+)(r_+-r_-)} - \frac{(2Mr_-\omega-am)^2}{(r-r_-)(r_+-r_-)}\\
&+ \left(r^2+2M(r+2M)\right)\omega^2\Biggr] R(r) = K_l\, R(r),
\end{aligned}
\end{equation}
where $K_l$ is the separation constant.

To show the hidden conformal symmetry of the nonextremal Kerr black hole \cite{Castro:2010fd},  we consider the low-energy limit $\omega M \ll 1$.  In this case,  Eq.~\eqref{eq:KGtheta} reduces to
\begin{equation}
\left[\frac{1}{\sin\theta}\partial_\theta(\sin\theta\partial_\theta) - \frac{m^2}{\sin^2\theta}\right] S(\theta) = -K_l\, S(\theta),
\end{equation}
which is the standard Laplace equation on the 2-sphere with $K_l=l(l+1)$,  and Eq.~\eqref{eq:KGr} in the near region $r \ll 1/\omega$ becomes
\begin{equation}
\begin{aligned}
&\left[\partial_r(\Delta\partial_r) + \frac{(2Mr_+\omega-am)^2}{(r-r_+)(r_+-r_-)} - \frac{(2Mr_-\omega-am)^2}{(r-r_-)(r_+-r_-)}\right] R(r)\\
&= K_l\, R(r),
\end{aligned}
\end{equation}
which can be viewed as the eigenvalue equation of the Casimir operators of the $SL(2,R)_L \times SL(2,R)_R$ conformal symmetry.  The conformal weights of the field $\Psi$ are
\begin{equation}
(h_L,h_R) = (l,l).
\end{equation}
The $2\pi$ periodic identification of the angular coordinate $\phi$ requires the ground state of the dual CFT to be at temperatures
\begin{equation}
T_L = \frac{r_++r_-}{4\pi a},\quad T_R = \frac{r_+-r_-}{4\pi a}.
\end{equation}
The conformal symmetry above has been supported by the black hole entropy and scattering amplitudes \cite{Castro:2010fd} and confirmed using the covariant phase space formalism \cite{Haco:2018ske}.

\section{Scalar radiation from a star orbiting in the near region}\label{sec:scalar}

In this section,  we view a star orbiting in the near region as a test particle and couple it to a massless scalar field.  The resulting scalar radiation at the horizon is computed using the gravity and the CFT analyses,  and two computations are shown to agree.  Finally,  we reattach the far region and compute the radiation power at the future null infinity.

\subsection{Gravity analysis}

Consider a star orbiting in the near region,  of which the coordinates are given by $(r_*,\theta_*,t_*,\phi_*)$.  The geodesic equations of generic Kerr black hole orbits are usually written as follows \cite{Misner:1973prb}:
\begin{equation}
\begin{aligned}
\Sigma^4\left(\frac{dr_*}{d\tau}\right)^2 & = R(r_*)\, ,& \Sigma^4\left(\frac{d\theta_*}{d\tau}\right)^2 & = \Theta(\theta_*)\, ,\\
\Sigma^2\frac{dt_*}{d\tau} & = T(r_*,\theta_*)\, ,& \Sigma^2\frac{d\phi_*}{d\tau} & = \Phi(r_*,\theta_*)\, ,
\end{aligned}
\end{equation}
where
\begin{equation}
\begin{aligned}
R(r) & \equiv \left[E(r^2+a^2)-aL\right]^2 - \Delta\left[(L-aE)^2+Q+r^2\right],\\
\Theta(\theta) & \equiv Q - \cos^2\theta\left[a^2(1-E^2)+\frac{L^2}{\sin^2\theta}\right],\\
T(r,\theta) & \equiv E\left[\frac{(r^2+a^2)^2}{\Delta}-a^2\sin^2\theta\right] + aL\left(1-\frac{r^2+a^2}{\Delta}\right),\\
\Phi(r,\theta) & \equiv aE\left(\frac{r^2+a^2}{\Delta}-1\right) + L\left(\frac{1}{\sin^2\theta}-\frac{a^2}{\Delta}\right).
\end{aligned}
\end{equation}
Here, $E$ is the energy,  $L$ is the $z$ component of the angular momentum, and $Q$ is the Carter constant.  All of them are conserved quantities.  $\tau$ is the proper time.  Note that $E$ and $L$ have been divided by $m_0$, and $Q$ has been divided by $m_0^2$, where $m_0$ is the rest mass of the star.  Hence, $E$ is a dimensionless quantity in these new units.  Reparametrize the geodesic with $\lambda(\tau) \equiv \int_0^\tau \frac{dx}{r(x)^2+a^2\cos^2\theta(x)}$, then the motion equations become \cite{Drasco:2003ky}
\begin{equation}
\begin{aligned}
\left(\frac{dr_*}{d\lambda}\right)^2 & = R(r_*)\, , & \left(\frac{d\theta_*}{d\lambda}\right)^2 & = \Theta(\theta_*)\, ,\\
\frac{dt_*}{d\lambda} & = T(r_*,\theta_*)\, , & \frac{d\phi_*}{d\lambda} & = \Phi(r_*,\theta_*)\, .
\end{aligned}
\end{equation}
Note that the $r$ and $\theta$ motions become periodic:
\begin{equation}
r_*(\lambda) = r_*(\lambda+n\Lambda_r),\ \theta_*(\lambda) = \theta_*(\lambda+n\Lambda_\theta),
\end{equation}
with
\begin{equation}
\Lambda_r = 2\int_{r_{min}}^{r_{max}} \frac{dr}{R(r)^{1/2}},\ \Lambda_\theta = 4\int_{\theta_{min}}^{\pi/2} \frac{d\theta}{\Theta(\theta)^{1/2}}.
\end{equation}
Since both $T(r,\theta)$ and $\Phi(r,\theta)$ are real and  can be written in the form $f(r)+g(\theta)$,  we can expand $T(r,\theta)$ and $\Phi(r,\theta)$ as
\begin{equation}
\begin{aligned}
T(r,\theta) & = \Gamma + \sum_{k=1}^{\infty} (T_k^\theta e^{-ik\Upsilon_\theta\lambda}+\text{c.c.}) + \sum_{n=1}^{\infty} (T_n^r e^{-in\Upsilon_r\lambda}+\text{c.c.})\, ,\\
\Phi(r,\theta) & = \Upsilon_\phi + \sum_{k=1}^{\infty} (\Phi_k^\theta e^{-ik\Upsilon_\theta\lambda}+\text{c.c.}) + \sum_{n=1}^{\infty} (\Phi_n^r e^{-in\Upsilon_r\lambda}+\text{c.c.})\, ,
\end{aligned}
\end{equation}
where
\begin{equation}
\Upsilon_r \equiv 2\pi/\Lambda_r,\ \Upsilon_\theta \equiv 2\pi/\Lambda_\theta.
\end{equation}
Thus,  we have
\begin{equation}\label{eq:tstar and phistar}
t_*(\lambda) = \Gamma\lambda + \Delta t(\lambda),\ \phi_*(\lambda) = \Upsilon_\phi\lambda + \Delta \phi(\lambda),
\end{equation}
where
\begin{equation}
\begin{aligned}
\Delta t(\lambda) & = \sum_{k=1}^{\infty} (\Delta t_k^\theta e^{-ik\Upsilon_\theta\lambda}+\text{c.c.}) + \sum_{n=1}^{\infty} (\Delta t_n^r e^{-in\Upsilon_r\lambda}+\text{c.c.}),\\
\Delta \phi(\lambda) & = \sum_{k=1}^{\infty} (\Delta \phi_k^\theta e^{-ik\Upsilon_\theta\lambda}+\text{c.c.}) + \sum_{n=1}^{\infty} (\Delta \phi_n^r e^{-in\Upsilon_r\lambda}+\text{c.c.}),
\end{aligned}
\end{equation}
and
\begin{equation}
\begin{aligned}
\Delta t_k^\theta & \equiv iT_k^\theta/(k\Upsilon_\theta), & \Delta t_n^r & \equiv iT_n^r/(n\Upsilon_r),\\
\Delta \phi_k^\theta & \equiv i\Phi_k^\theta/(k\Upsilon_\theta), & \Delta \phi_n^r & \equiv i\Phi_n^r/(n\Upsilon_r).
\end{aligned}
\end{equation}
In principle,  given $E$,  $L$, and $Q$,  we can compute the corresponding orbit analytically, as shown above.

Coupling the star to a massless scalar field $\Psi$,  the action of this system can be written as
\begin{equation}
S = \frac{1}{2} \int d^4x \sqrt{-g} \left[\left(\partial\Psi(x)\right)^2 - 8\pi\kappa\Psi(x)\mathcal{S}(x)\right],
\end{equation}
where $\kappa$ is a coupling constant, and
\begin{equation*}
\mathcal{S}(x) \equiv \int d\tau (-g)^{-1/2} \delta^{(4)}\left(x-x_*(\tau)\right).
\end{equation*}
Then,  the scalar wave equation becomes
\begin{equation}\label{eq:scalar}
\square\Psi = 4\pi\mathcal{T},
\end{equation}
where the source term is
\begin{equation}
\begin{aligned}\label{eq:source term}
\mathcal{T} & \equiv -\kappa \int d\tau (-g)^{-1/2} \delta^{(4)}\left(x-x_*(\tau)\right)\\
& = -\kappa \int d\lambda \frac{1}{\sin\theta} \delta^{(4)}\left(x-x_*(\lambda)\right).
\end{aligned}
\end{equation}

Using the mode expansions
\begin{equation}\label{eq:mode}
\begin{aligned}
\Psi =& \int d\omega e^{-i\omega t} \sum_{l,m} e^{im\phi}S_l(\theta)R_{lm\omega}(r),\\
\mathcal{T} =& \frac{1}{4\pi\Sigma^2} \int d\omega e^{-i\omega t} \sum_{l,m} e^{im\phi}S_l(\theta)T_{lm\omega}(r),
\end{aligned}
\end{equation}
Eq.~\eqref{eq:scalar} in the low-energy and the near-region limits can be separated into
\begin{equation}
\frac{1}{\sin\theta}\partial_\theta(\sin\theta\partial_\theta S_l) - \frac{m^2}{\sin^2\theta}S_l = -K_l\, S_l,
\end{equation}
and
\begin{equation}\label{eq:scalarr}
\begin{aligned}
&\partial_r(\Delta\partial_rR_{lm\omega})  + \frac{(2Mr_+\omega-am)^2}{(r-r_+)(r_+-r_-)}R_{lm\omega} - \frac{(2Mr_-\omega-am)^2}{(r-r_-)(r_+-r_-)}R_{lm\omega}\\
&= K_l\, R_{lm\omega} + T_{lm\omega}.
\end{aligned}
\end{equation}
Since $e^{im\phi}S_l(\theta)$ in Eq.~\eqref{eq:mode} are spherical harmonics,  we have
\begin{equation}
\begin{aligned}
T_{lm\omega} &= \frac{1}{4\pi^2} \int_0^\pi \sin\theta d\theta \int_0^{2\pi} d\phi \int_{-\infty}^\infty dt 4\pi\Sigma^2\mathcal{T} e^{i\omega t}e^{-im\phi}S_l(\theta)\\
&= -\frac{\kappa}{\pi} \int_{-\infty}^{\infty} d\lambda e^{i\omega\Gamma\lambda-im\Upsilon_\phi\lambda} \left(f_1g_1+f_2g_2\right),
\end{aligned}
\end{equation}
where in the second line we have applied Eq.~\eqref{eq:source term} and Eq.~\eqref{eq:tstar and phistar},  and
\begin{equation}
\begin{aligned}
f_1 & \equiv r_*^2(\lambda)\delta\left(r-r_*(\lambda)\right)\\
& \cdot \exp\left[i\omega\sum_{n=1}^{\infty}(\Delta t_n^r e^{-in\Upsilon_r\lambda}+\text{c.c.}) - im\sum_{n=1}^{\infty}(\Delta \phi_n^r e^{-in\Upsilon_r\lambda}+\text{c.c.})\right],\\
g_1 & \equiv S_l\left(\theta_*(\lambda)\right)\\
& \cdot \exp\left[i\omega\sum_{k=1}^{\infty}(\Delta t_k^\theta e^{-ik\Upsilon_\theta\lambda}+\text{c.c.}) - im\sum_{k=1}^{\infty}(\Delta \phi_k^\theta e^{-ik\Upsilon_\theta\lambda}+\text{c.c.})\right],\\
f_2 & \equiv \delta\left(r-r_*(\lambda)\right)\\
& \cdot \exp\left[i\omega\sum_{n=1}^{\infty}(\Delta t_n^r e^{-in\Upsilon_r\lambda}+\text{c.c.}) - im\sum_{n=1}^{\infty}(\Delta \phi_n^r e^{-in\Upsilon_r\lambda}+\text{c.c.})\right],\\
g_2 & \equiv a^2\cos^2\theta_*(\lambda)S_l\left(\theta_*(\lambda)\right)\\
& \cdot \exp\left[i\omega\sum_{k=1}^{\infty}(\Delta t_k^\theta e^{-ik\Upsilon_\theta\lambda}+\text{c.c.}) - im\sum_{k=1}^{\infty}(\Delta \phi_k^\theta e^{-ik\Upsilon_\theta\lambda}+\text{c.c.})\right].
\end{aligned} 
\end{equation}
Note that $T_{lm\omega}(r)$ is nonzero only when $r_{min} \leq r \leq r_{max}$ because of the existence of the term $\delta\left(r-r_*(\lambda)\right)$.  Since $f_1(\lambda+n\Lambda_r) = f_1(\lambda)$,  $g_1(\lambda+n\Lambda_\theta) = g_1(\lambda)$,  $f_2(\lambda+n\Lambda_r) = f_2(\lambda)$ and $g_2(\lambda+n\Lambda_\theta) = g_2(\lambda)$,  we can expand
\begin{equation}
f_1g_1+f_2g_2 = \sum_{k,n} f_{lm\omega,kn}(r) e^{-ik\Upsilon_\theta\lambda}e^{-in\Upsilon_r\lambda}.
\end{equation}
Thus,
\begin{equation}\label{eq:scalarT}
\begin{aligned}
T_{lm\omega} &= -\frac{\kappa}{\pi} \sum_{k,n} f_{lm\omega,kn}(r) \int_{-\infty}^\infty d\lambda e^{i\omega\Gamma\lambda-im\Upsilon_\phi\lambda}e^{-ik\Upsilon_\theta\lambda}e^{-in\Upsilon_r\lambda},\\
&= -\frac{2\kappa}{\Gamma} \sum_{k,n} f_{lm\omega,kn}(r)\, \delta(\omega-m\Omega_\phi-k\Omega_\theta-n\Omega_r),
\end{aligned}
\end{equation}
where
\begin{equation}
\Omega_\phi \equiv \Upsilon_\phi/\Gamma,\ \Omega_\theta \equiv \Upsilon_\theta/\Gamma,\ \Omega_r \equiv \Upsilon_r/\Gamma.
\end{equation}

Equation \eqref{eq:scalarr} is a Sturm-Liouville problem,  and we will show that it can be solved via the standard Green's function method for given boundary conditions.  Firstly,  two linearly independent solutions to the homogeneous equation (with $T_{lm\omega}=0$) can be written as
\begin{equation}
\begin{aligned}
R_{lm\omega}^{(1)} & = \left(\frac{r-r_+}{r-r_-}\right)^{-i\eta_1}\left(1-\frac{r-r_+}{r-r_-}\right)^{l+1}\\
 & \cdot\, _2F_1\left(1+l-i\eta_2,1+l-i2M\omega;1-2i\eta_1;\frac{r-r_+}{r-r_-}\right)\\
 & = \left(\frac{r-r_+}{r-r_-}\right)^{-i\eta_1}\left(1-\frac{r-r_+}{r-r_-}\right)^{l+1}\\
 & \cdot\, \Biggl[A\ _2F_1\left(1+l-i\eta_2,1+l-i2M\omega;2l+2;1-\frac{r-r_+}{r-r_-}\right)\\
 & + B\,\left(1-\frac{r-r_+}{r-r_-}\right)^{-2l-1}\\
 & \, \cdot \ _2F_1\left(-l-i2M\omega,-l-i\eta_2;-2l;1-\frac{r-r_+}{r-r_-}\right)\Biggr]\, ,
\end{aligned}
\end{equation}
\begin{equation}
\begin{aligned}
R_{lm\omega}^{(2)} & = \left(\frac{r-r_+}{r-r_-}\right)^{i\eta_1}\left(1-\frac{r-r_+}{r-r_-}\right)^{l+1}\\
 & \cdot\, _2F_1\left(1+l+i2M\omega,1+l+i\eta_2;1+2i\eta_1;\frac{r-r_+}{r-r_-}\right)\\
 & = \left(\frac{r-r_+}{r-r_-}\right)^{i\eta_1}\left(1-\frac{r-r_+}{r-r_-}\right)^{l+1}\\
 & \cdot\, \Biggl[C\ _2F_1\left(1+l+i2M\omega,1+l+i\eta_2;2l+2;1-\frac{r-r_+}{r-r_-}\right)\\
 & + D\, \left(1-\frac{r-r_+}{r-r_-}\right)^{-2l-1}\\
 & \,\cdot \ _2F_1\left(-l+i\eta_2,-l+i2M\omega;-2l;1-\frac{r-r_+}{r-r_-}\right)\Biggr],
\end{aligned}
\end{equation}
where $\ _2F_1(\alpha,\beta;\gamma;z)$ is the hypergeometric function,
\begin{equation}
\eta_1 \equiv \frac{2Mr_+\omega-am}{r_+-r_-},\ \eta_2 \equiv \frac{4M^2\omega-2am}{r_+-r_-},
\end{equation}
and
\begin{equation}
\begin{aligned}
A & \equiv \frac{\Gamma(1-2i\eta_1)\Gamma(-2l-1)}{\Gamma(-l-i2M\omega)\Gamma(-l-i\eta_2)},\\
B & \equiv \frac{\Gamma(1-2i\eta_1)\Gamma(2l+1)}{\Gamma(1+l-i\eta_2)\Gamma(1+l-i2M\omega)},\\
C & \equiv \frac{\Gamma(1+2i\eta_1)\Gamma(-2l-1)}{\Gamma(-l+i\eta_2)\Gamma(-l+i2M\omega)},\\
D & \equiv \frac{\Gamma(1+2i\eta_1)\Gamma(2l+1)}{\Gamma(1+l+i2M\omega)\Gamma(1+l+i\eta_2)}.
\end{aligned}
\end{equation}
Here,  we have used the following identity for convenience to obtain the large-$r$ behaviors of the solutions:
\begin{equation}
\begin{aligned}
&\ _2F_1(\alpha,\beta;\gamma;z)\\
= &\,  \frac{\Gamma(\gamma)\, \Gamma(\gamma-\alpha-\beta)}{\Gamma(\gamma-\alpha)\, \Gamma(\gamma-\beta)}\ _2F_1(\alpha,\beta;\alpha+\beta-\gamma+1;1-z)\\
 & +(1-z)^{\gamma-\alpha-\beta}\frac{\Gamma(\gamma)\, \Gamma(\alpha+\beta-\gamma)}{\Gamma(\alpha)\, \Gamma(\beta)}\\
 & \cdot \ _2F_1(\gamma-\alpha,\gamma-\beta;\gamma-\alpha-\beta+1;1-z).
\end{aligned}
\end{equation}
The asymptotic behaviors of the solutions are
\begin{equation}
\begin{aligned}
R_{lm\omega}^{(1)} &\to \left(\frac{r-r_+}{r_+-r_-}\right)^{-i\eta_1}&&r \to r_+,\\
&\to A(r_+-r_-)^{l+1}r^{-l-1} + B(r_+-r_-)^{-l}r^l&&r \gg M,
\end{aligned}
\end{equation}
\begin{equation}
\begin{aligned}
R_{lm\omega}^{(2)} &\to \left(\frac{r-r_+}{r_+-r_-}\right)^{i\eta_1}&&r \to r_+,\\
&\to C(r_+-r_-)^{l+1}r^{-l-1} + D(r_+-r_-)^{-l}r^l&&r \gg M.
\end{aligned}
\end{equation}
The large-$r$ behaviors of the solutions can be seen easily if we notice that $r \gg M$ is equivalent to $(r-r_+)/(r-r_-) \sim 1$.  As shown in \cite{Teukolsky:1973ha},  $R_{lm\omega}^{(1)}$ is purely ingoing at the horizon,  while $R_{lm\omega}^{(2)}$ is purely outgoing.  Both solutions have one rise and one falloff at the asymptotic boundary of the near region ($r \gg M$).  Let us define
\begin{equation}
R_{lm\omega}^{(3)} \equiv DR_{lm\omega}^{(1)} - BR_{lm\omega}^{(2)},
\end{equation}
whose asymptotic behaviors are
\begin{equation}
\begin{aligned}
R_{lm\omega}^{(3)} & \to D\left(\frac{r-r_+}{r_+-r_-}\right)^{-i\eta_1} - B\left(\frac{r-r_+}{r_+-r_-}\right)^{i\eta_1}&&r \to r_+,\\
& \to (DA-BC)(r_+-r_-)^{l+1}r^{-l-1}&&r \gg M.
\end{aligned}
\end{equation}
We see that $R_{lm\omega}^{(3)}$ has only one falloff when $r \gg M$,  which we will refer to as the Neumann boundary condition as in \cite{Porfyriadis:2014fja}.  In contrast, the Dirichlet boundary condition has only one rise when $r \gg M$.  Then,  the solution to Eq.~\eqref{eq:scalarr},  which is purely ingoing at the horizon and obeys the Neumann boundary condition when $r \gg M$,  is given by
\begin{equation}
\begin{aligned}
&R_{lm\omega}(r)\\
= & \frac{1}{W}\int_{r_{min}}^{r_{max}} dr_0 T_{lm\omega}(r_0)\\
& \cdot\left(R_{lm\omega}^{(3)}(r_0)\Theta(r_0-r)R_{lm\omega}^{(1)}(r) + R_{lm\omega}^{(1)}(r_0)\Theta(r-r_0)R_{lm\omega}^{(3)}(r)\right),
\end{aligned}
\end{equation}
where $W$ is the $r$-independent Wronskian
\begin{equation}
W = \Delta\left(R_{lm\omega}^{(1)}(r)\partial_rR_{lm\omega}^{(3)}(r) - R_{lm\omega}^{(3)}(r)\partial_rR_{lm\omega}^{(1)}(r)\right).
\end{equation}
As mentioned above,  $T_{lm\omega}(r)$ is nonzero only when $r_{min} \leq r \leq r_{max}$.  Thus,  we have
\begin{equation}
\begin{aligned}
\Psi(r \to r_+) = & \int d\omega e^{-i\omega t} \sum_{l,m} e^{im\phi}S_l(\theta)\frac{X}{W}\left(\frac{r-r_+}{r_+-r_-}\right)^{-i\eta_1},\\
\Psi(r \gg M) = & \int d\omega e^{-i\omega t} \sum_{l,m} e^{im\phi}S_l(\theta)\\
& \cdot \frac{Z}{W}(DA-BC)(r_+-r_-)^{l+1}r^{-l-1},
\end{aligned}
\end{equation}
where
\begin{equation}
\begin{aligned}
X & \equiv \int_{r_{min}}^{r_{max}} dr_0 T_{lm\omega}(r_0)R_{lm\omega}^{(3)}(r_0),\\
Z & \equiv \int_{r_{min}}^{r_{max}} dr_0 T_{lm\omega}(r_0)R_{lm\omega}^{(1)}(r_0).
\end{aligned}
\end{equation}

The Klein-Gordon particle number flux,  which is defined as
\begin{equation}
\mathcal{F} = \frac{1}{2\pi} \int \sqrt{-g} J^r d\theta d\phi dt,\ J^\mu = \frac{i}{8\pi}(\Psi^*\nabla^\mu\Psi - \Psi\nabla^\mu\Psi^*),
\end{equation}
vanishes at the asymptotic boundary of the near region as required by the Neumann boundary condition.  The flux at the horizon is
\begin{equation}\label{eq:scalarF}
\mathcal{F}_{lm\omega} = \frac{1}{2}(2Mr_+\omega-am)\frac{|X|^2}{|W|^2}.
\end{equation}
It may seem a little strange that the flux we get has such a simple form, considering the orbit of the star can be very complicated.  However,  that is because we put all the information of the orbit into $T_{lm\omega}$,  which is encoded in $X$ and $W$,  and the explicit expression of $T_{lm\omega}$ could be difficult to compute.

Finally,  let us discuss the validity of the low-energy limit $\omega M \ll 1$ and the near-region limit $\omega r \ll 1$ we used.  From Eq.~\eqref{eq:scalarT},  we know $\omega=m\Omega_\phi+k\Omega_\theta+n\Omega_r$,  where $\Omega_\phi$,  $\Omega_\theta$, and $\Omega_r$ are determined by the orbit conserved quantities $E$,  $L$, and $Q$.  It is difficult to precisely determine the parameter space, where both limits are satisfied simultaneously.  However,  if we focus on the star whose orbit radius $r_*$ is of the same order of magnitude as $M$ or just one order of magnitude higher than $M$,  numerical computations with typical parameter values support the validity of the approximations we used.  For example,  when $a=0.8M$,  $E=0.928$,  $L=2.10M$, and $Q=4.43M^2$,  we have $\Omega_\phi M=0.0679$,  $\Omega_\theta M=0.0615$, and $\Omega_r M=0.0356$ with $r_{max}=7M$,  $r_{min}=5M$, and $\theta_{min}=\pi/4$.  Another example is the retrograde orbit,  where $\Omega_\phi$ is negative,  rendering $\omega$ smaller.  For instance,  when $a=0.2M$,  $E=0.949$,  $L=-2.56M$, and $Q=6.58M^2$,  we have $\Omega_\phi M=-0.0490$,  $\Omega_\theta M=0.0500$, and $\Omega_r M=0.0184$ with $r_{max}=8M$,  $r_{min}=7M$, and $\theta_{min}=\pi/4$.  The details of computations of $\Omega_\phi$,  $\Omega_\theta$, and $\Omega_r$ for given $E$,  $L$, and $Q$ are shown in Appendix~\ref{app:omega}.  From the numerical results of several representative cases,  we see that the low-energy and the near-region limits are well satisfied in most regions relevant to black hole physics.

\subsection{CFT analysis}

According to the holographic dictionary, the Neumann modes satisfying the Neumann boundary condition correspond to the expectation values of the CFT operators, and the Dirichlet modes satisfying the Dirichlet boundary condition correspond to the external sources. Since we chose the Neumann boundary condition in the previous subsection, the dual CFT has no external sources or equivalent deformations, which seems strange because there is an orbiting star in the near region. In fact,  as analyzed in \cite{Porfyriadis:2014fja}, the orbiting star leads to the dual CFT deformed in the low-energy scale corresponding to the inside of the orbit. This effect can be reproduced by introducing appropriate external sources or equivalent deformations in the CFT. To compute the deformations explicitly,  we need to extend the solution in the region $r<r_{min}$ to the asymptotic boundary of the near region, where the extended solution will contain both the Neumann and the Dirichlet modes. The Dirichlet modes correspond to the effective deformations, so we can read off the CFT deformations from the coefficients of the Dirichlet modes.

The action of the deformed CFT is
\begin{equation}
S = S_{CFT} + \sum_l \int dt^+ dt^- J_l(t^+,t^-)\mathcal{O}_l(t^+,t^-),
\end{equation}
where $S_{CFT}$ is the original CFT action,  $J_l$ are the deformations, and $\mathcal{O}_l$ are the CFT operators with the left and the right conformal weights $(l, \, l)$.  A comparison between conformal coordinates of the Kerr metric \cite{Castro:2010fd} and those of $\mathrm{AdS}_3$ \cite{Maldacena:1998bw} implies
\begin{equation}
t^+ = \phi - \frac{a}{2M^2}t,\ t^- = -\phi.
\end{equation}
We expand the deformation $J_l$ as
\begin{equation}
J_l(t^+,t^-) = \int d\omega \sum_m J_{lm\omega} e^{i\omega_Lt^+ + i\omega_Rt^-},
\end{equation}
where
\begin{equation}
\omega_L = \frac{2M^2\omega}{a},\ \omega_R = \frac{2M^2\omega-am}{a}.
\end{equation}
As discussed above,  the coefficients $J_{lm\omega}$ are determined by the solution extended from $r < r_{min}$ to the asymptotic boundary of the near region:
\begin{equation}
\begin{aligned}
R^{ext}_{lm\omega}(r) &= \frac{X}{W}R_{lm\omega}^{(1)}(r)\\
&\to \frac{X}{W}\left(A(r_+-r_-)^{l+1}r^{-l-1} + B(r_+-r_-)^{-l}r^l\right),\ r \gg M.
\end{aligned}
\end{equation}
We read off $J_{lm\omega}$ from the Dirichlet modes:
\begin{equation}
J_{lm\omega} = \frac{X}{W}B.
\end{equation}
Thus,  the deformed action can be written as
\begin{equation}
S = S_{CFT} + \sum_{l,m} \int d\omega \int dt^+ dt^- J_{lm\omega}e^{i\omega_Lt^+ + i\omega_Rt^-}\mathcal{O}_l(t^+,t^-).
\end{equation}

The time-dependent perturbation pumps both the left and the right energies into the CFT.  Fermi's golden rule gives the transition rate out of the thermal state \cite{Maldacena:1997ih,  Gubser:1997cm}
\begin{equation}
\mathcal{R} = 2\pi \sum_{l,m} \int d\omega |J_{lm\omega}|^2\int dt^+ dt^- e^{-i\omega_Lt^+ - i\omega_Rt^-}G(t^+,t^-),
\end{equation}
where $G(t^+,t^-)=\langle\mathcal{O}^\dagger(t^+,t^-)\mathcal{O}(0,0)\rangle_{T_L,T_R}$ is the finite-temperature two-point function of the CFT.  Using $T_L = (r_++r_-)/4\pi a$ and $T_R = (r_+-r_-)/4\pi a$,  with the appropriate $i\epsilon$ prescription, we have \cite{Bredberg:2009pv,  Porfyriadis:2014fja,  Hadar:2014dpa,  Hadar:2015xpa}
\begin{equation}
\begin{aligned}
\mathcal{R}_{lm\omega} & = \frac{4\pi^2}{a} C_\mathcal{O}^2 \left(\frac{M}{a}\right)^{2l-1}\left(\frac{\sqrt{M^2-a^2}}{a}\right)^{2l-2} \frac{|X|^2}{|W|^2} (2Mr_+\omega-am)\\
 & \cdot \frac{1}{1-e^{-4\pi\frac{2Mr_+\omega-am}{r_+-r_-}}} \frac{4l^2}{(l^2+4M^2\omega^2)(l^2+(\frac{4M^2\omega-2am}{r_+-r_-})^2)}\\
 & \approx \frac{16\pi^2}{al^2} C_\mathcal{O}^2 \left(\frac{M}{a}\right)^{2l-1}\left(\frac{\sqrt{M^2-a^2}}{a}\right)^{2l-2} \frac{|X|^2}{|W|^2} (2Mr_+\omega-am)\\
 & \cdot \frac{1}{1-e^{-4\pi\frac{2Mr_+\omega-am}{r_+-r_-}}}\, ,
\end{aligned}
\end{equation}
where $C_\mathcal{O}^2$ is an undetermined operator normalization, and in the last step, we have dropped some subleading terms in the limit $\omega M \ll 1$.

\subsection{Comparison}

Unlike the extremal case,  the dual CFT we deal with here is at nonzero temperature.  When we compute the transition rate,  the effect of stimulated emission should be subtracted.  The net result is \cite{Gubser:1997cm}
\begin{equation}\label{eq:scalarRnet}
\mathcal{R}_{lm\omega}^{net} = \mathcal{R}_{lm\omega}(1-e^{-\beta_L\omega_L-\beta_R\omega_R}),
\end{equation}
where
\begin{equation}
\beta_L = 1/T_L,\quad \beta_R = 1/T_R.
\end{equation}
Comparing Eq.~\eqref{eq:scalarF} with Eq.~\eqref{eq:scalarRnet},  we see that by choosing the normalization
\begin{equation}
C_\mathcal{O} = \frac{\sqrt{a}l}{4\sqrt{2}\pi}\left(\frac{a}{M}\right)^{l-1/2}\left(\frac{a}{\sqrt{M^2-a^2}}\right)^{l-1},
\end{equation}
which is independent of $m$ and $\omega$,  we have exactly
\begin{equation}
\mathcal{R}_{lm\omega}^{net} = \mathcal{F}_{lm\omega},
\end{equation}
as predicted by the holographic dictionary.

Now,  let us discuss the extremal limit $a = M$.  In this limit,  it is tempting to suspect that the hidden conformal symmetry reduces to the conformal symmetry of the NHEK geometry.  However,  it should be emphasized that the near region for the nonextremal Kerr black hole is defined by $r-r_+ \ll 1/\omega$.  In contrast,  the near-horizon region of the extremal case describes the part $r-r_+ \ll M$,   which is stricter in the low-energy limit.  Thus,  the hidden conformal symmetry is not equal to the NHEK conformal symmetry even if $a = M$.

In fact,  the results obtained in this paper do not work well when taking $a \to M$ and $(r-r_+)/M \to 0$.  The modes that can propagate in the NHEK geometry are those with frequencies very close to the superradiant bound $\omega = m/2M$.  However,  the low-energy limit requires $\omega M \ll 1$,  which means $m \ll 1$ for the superradiant bound and contradicts the fact that $m$ is a positive integer.  Thus,  we do not expect that nonextremal results in this paper can be extended to the extremal limit directly.

\subsection{Reattaching the far region}

In the previous subsections,  we impose the Neumann boundary condition at the asymptotic boundary of the near region.  Here,  we reattach the far region and compute the scalar radiation flux at the future null infinity.

Using the mode expansions \eqref{eq:mode},  the radial part of Eq.~\eqref{eq:scalar} can be written as
\begin{equation}
\partial_r(\Delta\partial_rR_{lm\omega}) + \left(\frac{H^2}{\Delta}+2am\omega-a^2\omega^2\right)R_{lm\omega} = K_lR_{lm\omega} + T_{lm\omega},
\end{equation}
where $H=(r^2+a^2)\omega-am$.  Considering the low-energy limit $\omega M \ll 1$ and the far-region limit $r \gg M$,  we have
\begin{equation}\label{eq:scalarrf}
\partial_r(r^2\partial_rR_{lm\omega}) + \left(r^2\omega^2-l(l+1)\right)R_{lm\omega} = 0.
\end{equation}
Two linearly independent solutions are
\begin{equation}
\frac{1}{\sqrt{\omega r}}J_{l+1/2}(\omega r),\quad \frac{1}{\sqrt{\omega r}}J_{-l-1/2}(\omega r),
\end{equation}
with the asymptotic behaviors
\begin{equation}
\begin{aligned} 
\frac{1}{\sqrt{\omega r}}J_{l+1/2}(\omega r) &\to \omega^lr^l\frac{2^{-l-1/2}}{\Gamma(l+3/2)}&&r \ll 1/\omega,\\
&\to \sqrt{\frac{2}{\pi}}\frac{1}{\omega r}\sin(\omega r - \pi l/2)&&r \to \infty,
\end{aligned}
\end{equation}
\begin{equation}
\begin{aligned}
\frac{1}{\sqrt{\omega r}}J_{-l-1/2}(\omega r) &\to \omega^{-l-1}r^{-l-1}\frac{2^{l+1/2}}{\Gamma(-l+1/2)}&&r \ll 1/\omega,\\
&\to \sqrt{\frac{2}{\pi}}\frac{1}{\omega r}\cos(\omega r + \pi l/2)&&r \to \infty,
\end{aligned}
\end{equation}
where $J_\nu(z)$ are Bessel functions of the first kind.  Noticing that $J_\nu(\omega r) \sim \frac{1}{\Gamma(\nu+1)}\left(\frac{\omega r}{2}\right)^\nu$, when $\omega r \ll 1$, we can easily see the small-$r$ behaviors of the solutions.  The solution to Eq.~\eqref{eq:scalarrf}, which is purely outgoing at the null infinity, can be written as
\begin{equation}
R^{far}_{lm\omega}(r) = \frac{P}{\sqrt{\omega r}}J_{l+1/2}(\omega r) + \frac{Q}{\sqrt{\omega r}}J_{-l-1/2}(\omega r),
\end{equation}
where
\begin{equation}
\frac{P}{Q} = ie^{-i\pi l}.
\end{equation}
Since the far and the near regions overlap in the region $M \ll r \ll 1/\omega$,  we can match the inner part of the far-region solution with the outer part of the near-region solution.  The near-region solution that is purely ingoing at the horizon can be written as $R_{lm\omega}(r) + c_0R_{lm\omega}^{(1)}(r)$,  where $c_0$ is a constant.  We have
\begin{equation}
R^{far}_{lm\omega}(r \ll 1/\omega) = R_{lm\omega}(r \gg M) + c_0R_{lm\omega}^{(1)}(r \gg M).
\end{equation}
Then,  we can fix
\begin{equation}
\begin{aligned}
Q & = \frac{Z}{W}(DA-BC)(r_+-r_-)^{l+1}\\
  & \cdot \left[\omega^{-l-1}\frac{2^{l+1/2}}{\Gamma(-l+1/2)}-ie^{-i\pi l}\omega^l(r_+-r_-)^{2l+1}\frac{2^{-l-1/2}A}{\Gamma(l+3/2)B}\right]^{-1}.
\end{aligned}
\end{equation}
Thus,
\begin{equation}
\Psi(r \to \infty) = \int d\omega e^{-i\omega t} \sum_{l,m} e^{im\phi}S_l(\theta)Q\sqrt{\frac{2}{\pi}}\frac{1}{\omega r}e^{i\omega r}e^{i\pi l/2}.
\end{equation}

The outgoing radiation flux at the future null infinity is given by
\begin{equation}
\frac{dE}{dt} = \frac{1}{\pi}|Q|^2.
\end{equation}

\section{Gravitational radiation from a star orbiting in the near region}\label{sec:gravi}

In this section,  we generalize the analysis of the scalar radiation in the previous section to the gravitational case.

\subsection{Gravity analysis}

It is convenient to analyze gravitational waves in the Newman-Penrose formalism.  In Boyer-Lindquist coordinates,  we choose the Kinnersley tetrad \cite{Teukolsky:1973ha},  which has the components $(t,r,\theta,\phi)$ as
\begin{equation}
\begin{aligned}
l^\mu &= \left(\frac{r^2+a^2}{\Delta},1,0,\frac{a}{\Delta}\right),\\
n^\mu &= \frac{1}{2\Sigma^2}(r^2+a^2,-\Delta,0,a),\\
m^\mu &= \frac{1}{\sqrt{2}(r+ia\cos\theta)}\left(ia\sin\theta,0,1,\frac{i}{\sin\theta}\right).
\end{aligned}
\end{equation}
Note that $l \cdot n = -m \cdot \bar{m} = -1$.  In the standard notation,  the nonzero spin coefficients are
\begin{equation}
\begin{aligned}
\rho = -\frac{1}{r-ia\cos\theta},\ \tau = -\frac{ia\rho\bar{\rho}\sin\theta}{\sqrt{2}},\ \mu = \frac{(r^2-2Mr+a^2)\rho^2\bar{\rho}}{2},\\
\pi = \frac{ia\rho^2\sin\theta}{\sqrt{2}},\ \beta = -\frac{\bar{\rho}\cot\theta}{2\sqrt{2}},\ \gamma = \mu + \frac{(r-M)\rho\bar{\rho}}{2},\ \alpha = \pi - \bar{\beta},
\end{aligned}
\end{equation}
while the Weyl scalars are
\begin{equation}
\psi_0^{(0)} = \psi_1^{(0)} = \psi_3^{(0)} = \psi_4^{(0)} = 0,\ \psi_2^{(0)} = M\rho^3.
\end{equation}

Teukolsky showed \cite{Teukolsky:1973ha} that the perturbation $\psi_4^{(1)}$ of the Weyl scalar $\psi_4^{(0)}$ in any type D background (including the Schwarzschild and Kerr black holes) obeys
\begin{equation}\label{eq:gravi}
\begin{aligned}
&\Bigl[(\Delta+3\gamma-\bar{\gamma}+4\mu+\bar{\mu})(D+4\epsilon-\rho)\\
&-(\bar{\delta}-\bar{\tau}+\bar{\beta}+3\alpha+4\pi)(\delta-\tau+4\beta)-3\psi_2^{(0)}\Bigr]\psi_4^{(1)} = 4\pi T_4,
\end{aligned}
\end{equation}
where
\begin{equation}
\begin{aligned}
T_4 & = (\Delta+3\gamma-\bar{\gamma}+4\mu+\bar{\mu})\\
 & \cdot \left[(\bar{\delta}-2\bar{\tau}+2\alpha)T_{n\bar{m}}-(\Delta+2\gamma-2\bar{\gamma}+\bar{\mu})T_{\bar{m}\bar{m}}\right]\\
 & + (\bar{\delta}-\bar{\tau}+\bar{\beta}+3\alpha+4\pi)\\
 & \cdot \left[(\Delta+2\gamma+2\bar{\mu})T_{n\bar{m}}-(\bar{\delta}-\bar{\tau}+2\bar{\beta}+2\alpha)T_{nn}\right].
\end{aligned}
\end{equation}
Here,  $D \equiv l^\mu\partial_\mu$,  $\Delta \equiv n^\mu\partial_\mu$,  $\delta \equiv m^\mu\partial_\mu$,  and $\bar{\delta} \equiv \bar{m}^\mu\partial_\mu$.  The source terms are defined as
\begin{equation}
T_{n\bar{m}} \equiv T_{\mu\nu} n^\mu \bar{m}^\nu,\ T_{\bar{m}\bar{m}} \equiv T_{\mu\nu} \bar{m}^\mu \bar{m}^\nu,\ T_{nn} \equiv T_{\mu\nu} n^\mu n^\nu,
\end{equation}
where the energy-momentum tensor for a star of the rest mass $m_0$ is given by
\begin{equation}
\begin{aligned}
T^{\mu\nu} & = m_0 \int d\tau (-g)^{-1/2} \frac{dx^\mu_*}{d\tau}\frac{dx^\nu_*}{d\tau} \delta^{(4)}\left(x^\alpha-x_*^\alpha(\tau)\right)\\
& = m_0 \int d\lambda \frac{1}{\Sigma^4\sin\theta} \frac{dx^\mu_*}{d\lambda}\frac{dx^\nu_*}{d\lambda} \delta^{(4)}\left(x^\alpha-x_*^\alpha(\lambda)\right).
\end{aligned}
\end{equation}

To simplify Eq.~\eqref{eq:gravi},  we define
\begin{equation}
\psi^{(-2)} \equiv \rho^{-4}\psi_4^{(1)},\quad \mathcal{T} \equiv 2\rho^{-4}T_4,
\end{equation}
and use the expansions
\begin{equation}
\begin{aligned}
\psi^{(-2)} =& \int d\omega e^{-i\omega t} \sum_{l,m} e^{im\phi}S_l(\theta)R_{lm\omega}(r),\\
\mathcal{T} =& -\frac{1}{4\pi\Sigma^2} \int d\omega e^{-i\omega t} \sum_{l,m} e^{im\phi}S_l(\theta)T_{lm\omega}(r).
\end{aligned}
\end{equation}
Then,  Eq.~\eqref{eq:gravi} can be separated into
\begin{equation}\label{eq:gravitheta}
\begin{aligned}
&\frac{1}{\sin\theta}\partial_\theta(\sin\theta\partial_\theta S_l) - \frac{m^2}{\sin^2\theta}S_l - \frac{2ms\cos\theta}{\sin^2\theta}S_l - s^2\cot^2\theta S_l\\
&+sS_l - 2sa\omega\cos\theta S_l + a^2\omega^2\cos^2\theta S_l = -K_lS_l\, ,
\end{aligned}
\end{equation}
and
\begin{equation}\label{eq:gravir}
\begin{aligned}
&\Delta^{-s}\partial_r(\Delta^{s+1}\partial_rR_{lm\omega}) + \Biggl[\frac{(2Mr_+\omega-am)^2}{(r-r_+)(r_+-r_-)} - \frac{(2Mr_-\omega-am)^2}{(r-r_-)(r_+-r_-)}\\
&-is\frac{2M\omega r_+-am}{r-r_+} - is\frac{2M\omega r_--am}{r-r_-} + 2is\omega(r-M)\\
&+\left(r^2+2M(r+2M)\right)\omega^2\Biggr]R_{lm\omega} = K_lR_{lm\omega} + T_{lm\omega}\, ,
\end{aligned}
\end{equation}
where $K_l$ is the separation constant and $s = -2$,   compatible with the choice of the Weyl scalar $\psi_4$ as the outgoing wave \cite{Teukolsky:1972my}.  In the low-energy limit,  Eq.~\eqref{eq:gravitheta} becomes
\begin{equation}
\begin{aligned}
&\frac{1}{\sin\theta}\partial_\theta(\sin\theta\partial_\theta S_l) - \frac{m^2}{\sin^2\theta}S_l - \frac{2ms\cos\theta}{\sin^2\theta}S_l - s^2\cot^2\theta S_l\\
&+sS_l = -K_lS_l.
\end{aligned}
\end{equation}
Thus,  $e^{im\phi}S_l$ are the spin-weighted spherical harmonics with $K_l=(l-s)(l+s+1)$ \cite{Goldberg:1966uu},  and we can obtain
\begin{equation}
\begin{aligned}
&T_{lm\omega}(r)\\
=& \frac{1}{4\pi^2} \int_0^\pi \sin\theta d\theta \int_0^{2\pi} d\phi \int_{-\infty}^\infty dt e^{i\omega t}e^{-im\phi}S_l(\theta)(-8\pi\Sigma^2\rho^{-4}T_4)\\
=& \int_{r_{min}}^{r_{max}} dr_0 \bigl[a_0(r_0)\delta(r-r_0) + a_1(r_0)\delta'(r-r_0)\\
&+a_2(r_0)\delta''(r-r_0)\bigr],
\end{aligned}
\end{equation}
where the expressions of $a_0$,  $a_1$,  and $a_2$ are omitted,  whose explicit forms are irrelevant in the rest of this paper.  In the low-energy and the near-region limits,  Eq.~\eqref{eq:gravir} reduces to
\begin{equation}\label{eq:gravir1}
\begin{aligned}
&\Delta^{-s}\partial_r(\Delta^{s+1}\partial_rR_{lm\omega}) + \Biggl[\frac{(2Mr_+\omega-am)^2}{(r-r_+)(r_+-r_-)} - \frac{(2Mr_-\omega-am)^2}{(r-r_-)(r_+-r_-)}\\
&-is\frac{2M\omega r_+-am}{r-r_+} - is\frac{2M\omega r_--am}{r-r_-}\Biggr]R_{lm\omega} = K_lR_{lm\omega} + T_{lm\omega}.
\end{aligned}
\end{equation}

Equation \eqref{eq:gravir1} is also a Sturm-Liouville problem,  which can be solved via the standard Green's function method.  Two linearly independent solutions to the homogeneous equation ($T_{lm\omega}=0$) are given by hypergeometric functions
\begin{equation}
\begin{aligned}
&R_{lm\omega}^{(1)} = \left(\frac{r-r_+}{r-r_-}\right)^{-i\eta_1-s}\left(1-\frac{r-r_+}{r-r_-}\right)^{l+s+1}\\
& \cdot \ _2F_1\left(1+l-i\eta_2,1+l-s-i2M\omega;1-s-2i\eta_1;\frac{r-r_+}{r-r_-}\right)\\
& = \left(\frac{r-r_+}{r-r_-}\right)^{-i\eta_1-s}\left(1-\frac{r-r_+}{r-r_-}\right)^{l+s+1}\\
& \cdot \Biggl[A\ _2F_1\left(1+l-i\eta_2,1+l-s-i2M\omega;2l+2;1-\frac{r-r_+}{r-r_-}\right)\\
& + B\left(1-\frac{r-r_+}{r-r_-}\right)^{-2l-1}\\
& \cdot \ _2F_1\left(-l-s-i2M\omega,-l-i\eta_2;-2l;1-\frac{r-r_+}{r-r_-}\right)\Biggr],
\end{aligned}
\end{equation}
\begin{equation}
\begin{aligned}
&R_{lm\omega}^{(2)} = \left(\frac{r-r_+}{r-r_-}\right)^{i\eta_1}\left(1-\frac{r-r_+}{r-r_-}\right)^{l+s+1}\\
& \cdot \ _2F_1\left(1+l+s+i2M\omega,1+l+i\eta_2;1+s+2i\eta_1;\frac{r-r_+}{r-r_-}\right)\\
& = \left(\frac{r-r_+}{r-r_-}\right)^{i\eta_1}\left(1-\frac{r-r_+}{r-r_-}\right)^{l+s+1}\\
& \cdot \Biggl[C\ _2F_1\left(1+l+s+i2M\omega,1+l+i\eta_2;2l+2;1-\frac{r-r_+}{r-r_-}\right)\\
& + D\left(1-\frac{r-r_+}{r-r_-}\right)^{-2l-1}\\
& \cdot \ _2F_1\left(-l+i\eta_2,-l+s+i2M\omega;-2l;1-\frac{r-r_+}{r-r_-}\right)\Biggr],
\end{aligned}
\end{equation}
where
\begin{equation}
\eta_1 = \frac{2Mr_+\omega-am}{r_+-r_-},\ \eta_2 = \frac{4M^2\omega-2am}{r_+-r_-},
\end{equation}
and
\begin{equation}
\begin{aligned}
A &= \frac{\Gamma(1-s-2i\eta_1)\Gamma(-2l-1)}{\Gamma(-l-s-i2M\omega)\Gamma(-l-i\eta_2)},\\
B &= \frac{\Gamma(1-s-2i\eta_1)\Gamma(2l+1)}{\Gamma(1+l-i\eta_2)\Gamma(1+l-s-i2M\omega)},\\
C &= \frac{\Gamma(1+s+2i\eta_1)\Gamma(-2l-1)}{\Gamma(-l+i\eta_2)\Gamma(-l+s+i2M\omega)},\\
D &= \frac{\Gamma(1+s+2i\eta_1)\Gamma(2l+1)}{\Gamma(1+l+s+i2M\omega)\Gamma(1+l+i\eta_2)}.
\end{aligned}
\end{equation}
Then, the asymptotic behaviors of the solutions are
\begin{equation}
\begin{aligned}
R_{lm\omega}^{(1)} &\to \left(\frac{r-r_+}{r_+-r_-}\right)^{-i\eta_1-s}&&r \to r_+,\\
&\to A(r_+-r_-)^{l+s+1}r^{-l-s-1} + B(r_+-r_-)^{-l+s}r^{l-s}&&r \gg M,
\end{aligned}
\end{equation}
\begin{equation}
\begin{aligned}
R_{lm\omega}^{(2)} &\to \left(\frac{r-r_+}{r_+-r_-}\right)^{i\eta_1}&&r \to r_+,\\
&\to C(r_+-r_-)^{l+s+1}r^{-l-s-1} + D(r_+-r_-)^{-l+s}r^{l-s}&&r \gg M.
\end{aligned}
\end{equation}
Like the scalar field case,  we define
\begin{equation}
R_{lm\omega}^{(3)} \equiv DR_{lm\omega}^{(1)} - BR_{lm\omega}^{(2)},
\end{equation}
whose asymptotic behaviors are
\begin{equation}
\begin{aligned}
R_{lm\omega}^{(3)} &\to D\left(\frac{r-r_+}{r_+-r_-}\right)^{-i\eta_1-s} - B\left(\frac{r-r_+}{r_+-r_-}\right)^{i\eta_1}&&r \to r_+,\\
&\to (DA-BC)(r_+-r_-)^{l+s+1}r^{-l-s-1}&&r \gg M.
\end{aligned}
\end{equation}
The solution to Eq.~\eqref{eq:gravir1},  which is purely ingoing at the horizon and obeys the Neumann boundary conditions when $r \gg M$,  is given by
\begin{equation}
\begin{aligned}
R_{lm\omega}(r) =& \int_{r_{min}}^{r_{max}} dr_0 \biggl[\frac{1}{\Delta_0^{-s}W}\Bigl(X(r_0)\Theta(r_0-r)R_{lm\omega}^{(1)}(r)\\
&+ Z(r_0)\Theta(r-r_0)R_{lm\omega}^{(3)}(r)\Bigr) + \frac{a_2(r_0)}{\Delta_0}\delta(r-r_0)\biggr],
\end{aligned}
\end{equation}
where $W$ is the $r$-independent Wronskian
\begin{equation}
W = \Delta^{s+1}\left(R_{lm\omega}^{(1)}(r)\partial_rR_{lm\omega}^{(3)}(r) - R_{lm\omega}^{(3)}(r)\partial_rR_{lm\omega}^{(1)}(r)\right),
\end{equation}
and
\begin{equation}
\begin{aligned}
X(r_0) & = R_{lm\omega}^{(3)}(r_0)\biggl(a_0(r_0) - s\frac{(2r_0-2M)a_1(r_0)}{\Delta_0}\\
&+s(s-1)\frac{(2r_0-2M)^2a_2(r_0)}{\Delta_0^2} + (2s-V(r_0))\frac{a_2(r_0)}{\Delta_0}\biggr)\\
&+\partial_rR_{lm\omega}^{(3)}(r_0)\left(-a_1(r_0) + (s-1)\frac{(2r_0-2M)a_2(r_0)}{\Delta_0}\right),\\
Z(r_0) & = R_{lm\omega}^{(1)}(r_0)\biggl(a_0(r_0) - s\frac{(2r_0-2M)a_1(r_0)}{\Delta_0}\\
&+s(s-1)\frac{(2r_0-2M)^2a_2(r_0)}{\Delta_0^2} + (2s-V(r_0))\frac{a_2(r_0)}{\Delta_0}\biggr)\\
&+\partial_rR_{lm\omega}^{(1)}(r_0)\left(-a_1(r_0) + (s-1)\frac{(2r_0-2M)a_2(r_0)}{\Delta_0}\right).
\end{aligned}
\end{equation}
Here, 
\begin{equation}
\Delta_0 = r_0^2 - 2Mr_0 + a^2,
\end{equation}
and
\begin{equation}
\begin{aligned}
V(r) & = \frac{(2Mr_+\omega-am)^2}{(r-r_+)(r_+-r_-)} - \frac{(2Mr_-\omega-am)^2}{(r-r_-)(r_+-r_-)} - is\frac{2M\omega r_+-am}{r-r_+}\\
&-is\frac{2M\omega r_--am}{r-r_-} - (l-s)(l+s+1).
\end{aligned}
\end{equation}
Thus,  we have
\begin{equation}
\begin{aligned}
\psi^{(-2)}(r \to r_+) =& \int d\omega e^{-i\omega t} \sum_{l,m} e^{im\phi}S_l(\theta)\, \mathcal{X}\left(\frac{r-r_+}{r_+-r_-}\right)^{-i\eta_1-s},\\
\psi^{(-2)}(r \gg M) =& \int d\omega e^{-i\omega t} \sum_{l,m} e^{im\phi}S_l(\theta)\\
& \cdot \mathcal{Z}(DA-BC)(r_+-r_-)^{l+s+1}r^{-l-s-1},
\end{aligned}
\end{equation}
where
\begin{equation}
\mathcal{X} = \int_{r_{min}}^{r_{max}} dr_0 \frac{X(r_0)}{\Delta_0^{-s}W},\ \mathcal{Z} = \int_{r_{min}}^{r_{max}} dr_0 \frac{Z(r_0)}{\Delta_0^{-s}W}.
\end{equation}

The equivalent particle number flux at the horizon,  which has been studied in \cite{Teukolsky:1974yv},  is given by
\begin{equation}
\begin{aligned}
\mathcal{F}_{lm\omega} & = \frac{128(2Mr_+\omega-am)}{|\mathcal{C}|^2(r_+-r_-)^4}|\mathcal{X}|^2\\
& \cdot \left(\frac{(2Mr_+\omega-am)^2}{(r_+-r_-)^2}+\frac{1}{4}\right)\left(\frac{(2Mr_+\omega-am)^2}{(r_+-r_-)^2}+1\right),
\end{aligned}
\end{equation}
where
\begin{equation}
\begin{aligned}
|\mathcal{C}|^2 & = (Q^2 + 4a\omega m - 4a^2\omega^2)[(Q-2)^2 + 36a\omega m - 36a^2\omega^2]\\
& +(2Q-1)(96a^2\omega^2 - 48a\omega m) + 144\omega^2(M^2-a^2),\\
Q & = l^2 + l + a^2\omega^2 - 2a\omega m.
\end{aligned}
\end{equation}

\subsection{CFT analysis and comparison}

Like the scalar case,  the orbiting star leads to a deformation of the dual CFT and the action of the deformed CFT can be written as
\begin{equation}
S = S_{CFT} + \sum_{l,m} \int d\omega \int dt^+ dt^- J_{lm\omega}e^{i\omega_Lt^+ + i\omega_Rt^-}\mathcal{O}_l(t^+,t^-).
\end{equation}
Here,  for the gravitational field,  the conformal weights of the operator $\mathcal{O}_l$ are $(l-s,l)$ \cite{Hartman:2009nz}.  The deformation coefficients $J_{lm\omega}$ are determined by the metric perturbation,  which can be given by a Hertz potential $\Psi_H$ as follows \cite{Hartman:2009nz}:
\begin{equation}
h_{\mu\nu} = \Theta_{\mu\nu}\Psi_H,
\end{equation}
where
\begin{equation}
\begin{aligned}
\Theta_{\mu\nu} =& -l_\mu l_\nu(\bar{\delta}+\alpha+3\bar{\beta}-\bar{\tau})(\bar{\delta}+4\bar{\beta}+3\bar{\tau})\\
&-{\bar{m}}_\mu{\bar{m}}_\nu(D-\bar{\rho})(D+3\bar{\rho})\\
&+l_{(\mu}{\bar{m}}_{\nu)}\Bigl[(D+\rho-\bar{\rho})(\bar{\delta}+4\bar{\beta}+3\bar{\tau})\\
&+(\bar{\delta}+3\bar{\beta}-\alpha-\pi-\bar{\tau})(D+3\bar{\rho})\Bigr]
\end{aligned}
\end{equation}
and
\begin{equation}
\Psi_H = \frac{4}{\mathcal{C}}R^{(-2)}(r)S^{(+2)}(\theta)e^{-i\omega t+im\phi}.
\end{equation}
To get the deformation,  we extend the solution in the region $r<r_{min}$ to the asymptotic boundary of the near region,  so
\begin{equation}
\begin{aligned}
& R^{(-2)}(r)\\
& = \mathcal{X}R_{lm\omega}^{(1)}(r)\\
& \to \mathcal{X}[A(r_+-r_-)^{l+s+1}r^{-l-s-1} + B(r_+-r_-)^{-l+s}r^{l-s}],&&r \gg M.
\end{aligned}
\end{equation}
As analyzed in \cite{Porfyriadis:2014fja},  the coefficients $J_{lm\omega}$ are read off from the Dirichlet modes of the Hertz potential:
\begin{equation}
J_{lm\omega} = \frac{4}{\mathcal{C}}\mathcal{X}B.
\end{equation}

Following the steps similar to the scalar case,  we obtain
\begin{equation}
\begin{aligned}
&\mathcal{R}_{lm\omega}\\
& = 128\pi^2 C_\mathcal{O}^2 \left(\frac{M}{a}\right)^{2l+3}\left(\frac{\sqrt{M^2-a^2}}{a}\right)^{2l-1} |\mathcal{X}|^2 \frac{2Mr_+\omega-am}{|\mathcal{C}|^2(r_+-r_-)}\\
& \cdot \frac{1}{1-e^{-4\pi\frac{2Mr_+\omega-am}{r_+-r_-}}}\frac{2l}{(2l+1)(2l+2)(2l+3)}\\
& \cdot \frac{(4+4(\frac{2Mr_+\omega-am}{r_+-r_-})^2)(1+4(\frac{2Mr_+\omega-am}{r_+-r_-})^2)}{((l+2)^2+(2M\omega)^2)(l^2+(\frac{4M^2\omega-2am}{r_+-r_-})^2)}\\
& \approx \frac{2^{15}a^3\pi^2 C_\mathcal{O}^2}{l(2l+1)(2l+2)(2l+3)(l+2)^2} \left(\frac{M}{a}\right)^{2l+3}\left(\frac{\sqrt{M^2-a^2}}{a}\right)^{2l+2}\\
& \cdot \frac{2Mr_+\omega-am}{|\mathcal{C}|^2(r_+-r_-)^4}|\mathcal{X}|^2 \left(\frac{(2Mr_+\omega-am)^2}{(r_+-r_-)^2}+\frac{1}{4}\right)\\
& \cdot \left(\frac{(2Mr_+\omega-am)^2}{(r_+-r_-)^2}+1\right) \frac{1}{1-e^{-4\pi\frac{2Mr_+\omega-am}{r_+-r_-}}}\, .
\end{aligned}
\end{equation}
In the last step,  we dropped some subleading terms in the limit $\omega M \ll 1$.

With the normalization
\begin{equation}
\begin{aligned}
C_\mathcal{O} =& \frac{\sqrt{l(2l+1)(2l+2)(2l+3)}(l+2)}{16a^{3/2}\pi}\\
& \cdot \left(\frac{a}{M}\right)^{l+3/2} \left(\frac{a}{\sqrt{M^2-a^2}}\right)^{l+1},
\end{aligned}
\end{equation}
we find again
\begin{equation}
\mathcal{R}_{lm\omega}(1-e^{-\beta_L\omega_L-\beta_R\omega_R}) = \mathcal{F}_{lm\omega}.
\end{equation}

\subsection{Reattaching the far region}

In this subsection,  we reattach the far region and compute the gravitational radiation flux at the future null infinity.

Equation \eqref{eq:gravir} can be written as
\begin{equation}
\begin{aligned}
&\Delta^{-s}\partial_r(\Delta^{s+1}\partial_rR_{lm\omega}) + \Biggl[\frac{H^2-2is(r-M)H}{\Delta} + 4is\omega r\\
&+2am\omega - a^2\omega^2\Biggr] R_{lm\omega} = K_lR_{lm\omega} + T_{lm\omega},
\end{aligned}
\end{equation}
where $H=(r^2+a^2)\omega-am$.  Considering the low-energy and the far-region limits,  we have
\begin{equation}\label{eq:gravirf}
\begin{aligned}
&r^{-2s}\partial_r(r^{2s+2}\partial_rR_{lm\omega})\\
&+\left(r^2\omega^2 + 2is\omega r - (l-s)(l+s+1)\right)R_{lm\omega} = 0.
\end{aligned}
\end{equation}
Two linearly independent solutions are given by confluent hypergeometric functions
\begin{equation}
\begin{aligned}
R_{lm\omega}^{(1,far)} & = r^{l-s}e^{-i\omega r}\ _1F_1(l+1-s,2l+2;2i\omega r),\\
R_{lm\omega}^{(2,far)} & = r^{-l-s-1}e^{-i\omega r}\ _1F_1(-l-s,-2l;2i\omega r),
\end{aligned}
\end{equation}
with the asymptotic behaviors
\begin{equation}
\begin{aligned}
R_{lm\omega}^{(1,far)} & \to r^{l-s},&&r \ll 1/\omega,\\
& \to \frac{(-2i\omega)^{-1-l+s}\Gamma(2+2l)}{\Gamma(1+l+s)}r^{-1}e^{-i\omega r}\\
& +\frac{(2i\omega)^{-1-l-s}\Gamma(2+2l)}{\Gamma(1+l-s)}r^{-1-2s}e^{i\omega r},&&r \to \infty,
\end{aligned}
\end{equation}
\begin{equation}
\begin{aligned}
R_{lm\omega}^{(2,far)} & \to r^{-l-s-1},&&r \ll 1/\omega,\\
& \to \frac{(-2i\omega)^{l+s}\Gamma(-2l)}{\Gamma(-l+s)}r^{-1}e^{-i\omega r}\\
& +\frac{(2i\omega)^{l-s}\Gamma(-2l)}{\Gamma(-l-s)}r^{-1-2s}e^{i\omega r},&&r \to \infty.
\end{aligned}
\end{equation}
The solution to Eq.~\eqref{eq:gravirf}, which is purely outgoing at the future null infinity, can be written as
\begin{equation}
R_{lm\omega}^{(far)}(r) = PR_{lm\omega}^{(1,far)} + QR_{lm\omega}^{(2,far)},
\end{equation}
where
\begin{equation}
P/Q = -(-2i\omega)^{2l+1}\frac{\Gamma(1+l+s)\Gamma(-2l)}{\Gamma(-l+s)\Gamma(2l+2)}.
\end{equation}
Matching the far-region solution with the near-region solution,  we obtain
\begin{equation}
\begin{aligned}
Q & = \mathcal{Z}(DA-BC)(r_+-r_-)^{l+s+1}\\
& \cdot \left[1+(-2i\omega)^{2l+1}\frac{\Gamma(1+l+s)\Gamma(-2l)}{\Gamma(-l+s)\Gamma(2l+2)}\frac{A}{B}(r_+-r_-)^{2l+1}\right]^{-1}.
\end{aligned}
\end{equation}
Thus,
\begin{equation}
\begin{aligned}
\psi^{(-2)}(r \to \infty) & = \int d\omega e^{-i\omega t} \sum_{l,m} e^{im\phi}S_l(\theta)Q\frac{(2i\omega)^{l-s}\Gamma(-2l)}{\Gamma(-l-s)}\\
& \cdot \left[1+\frac{\Gamma(-l-s)\Gamma(1+l+s)}{\Gamma(-l+s)\Gamma(1+l-s)}\right]r^{-1-2s}e^{i\omega r}.
\end{aligned}
\end{equation}

The outgoing radiation flux at the future null infinity,  which has been studied in \cite{Teukolsky:1974yv},  is given by
\begin{equation}
\frac{dE}{dt} = 2^{2l+5}\omega^{2l+2}|Q|^2\left|\frac{\Gamma(-2l)}{\Gamma(-l+2)}\right|^2.
\end{equation}

\section{Discussion}\label{sec:discu}

In this paper,  we compute gravitational waves emitted by a star orbiting near the horizon of a nonextremal Kerr black hole from the gravity and the dual CFT perspectives,  respectively.  Considering the stimulated emission at finite temperature,  two results agree.  This strongly supports the hidden conformal symmetry of the Kerr black hole.  The stimulated emission factor at finite temperature is notable.  In previous studies,  since the extreme Kerr black hole is dual to a CFT at zero temperature,  this factor does not appear explicitly in the results.

To show the hidden conformal symmetry,  we take two limits in this paper.  The near-region limit is easy to understand.  When we are far enough away from the black hole,  the effect of the black hole is negligible.  Therefore,  the dual CFT mainly contains information about the black hole and the spacetime near it.  In other words,  the black hole and its near region are dual to the CFT.  The low-energy limit seems unnatural.  However, suppose we are close enough to a black hole,  and its effect on us is large enough.  In that case,  all physical activities (such as the generation and propagation of gravitational waves) are in low energy compared to the effect of the black hole.  Thus,  both limits imply the same condition: the black hole dominates the physical phenomena we consider.

Although we succeeded in computing gravitational waves using the hidden conformal symmetry in this paper,  there are still many difficulties in developing a useful CFT tool to compute gravitational waves in practice.  Firstly,  for a generic nonextremal Kerr black hole,  the conformal symmetry of the spacetime is destroyed,  so relating different orbits like the extremal case is not easy.  The solution space where the conformal symmetry exists might be a potential direction,  and various gravitational waves might be related through conformal transformations.  In addition,  the gravitational wave sources in astronomical observations are not necessarily limited to a test object orbiting near the horizon of a black hole.  Extending the application of CFT to more cases, such as the binary black hole merger,  requires a complete understanding of the Kerr/CFT correspondence.

\setcounter{secnumdepth}{0}

\section{acknowledgments}

We thank Jiang Long,  Jianxin Lu,  Wei Song,  and Yueliang Wu for helpful discussions.  This work is supported by the National Key Research and Development Program of China (Grant No.~2021YFC2201901) and in part by the NSFC under 
Grant No.~12147103.

\setcounter{secnumdepth}{2}

\appendix

\section{COMPUTATIONS OF $\Omega_\phi$,  $\Omega_\theta$, AND $\Omega_r$}\label{app:omega}
In this appendix,  we will show how to compute $\Omega_\phi$,  $\Omega_\theta$, and $\Omega_r$.

For given $E$,  $L$, and $Q$,  $r_{max}$ and $r_{min}$ are the roots to the equation $R(r)=0$, while $\theta_{min}$ is the root to the equation $\Theta(\theta)=0$.  If $R(r)$ is positive when $r_{min}<r<r_{max}$, and $\Theta(\theta)$ is positive when $\theta_{min}<\theta<\pi/2$,  then $\Lambda_r$ and $\Lambda_\theta$ are given by
\begin{equation}
\Lambda_r = 2\int_{r_{min}}^{r_{max}} \frac{dr}{R(r)^{1/2}},\quad \Lambda_\theta = 4\int_{\theta_{min}}^{\pi/2} \frac{d\theta}{\Theta(\theta)^{1/2}}.
\end{equation}
Note that $T(r,\theta)$ can be written as
\begin{equation}
T(r,\theta) = T_1(r) + T_2(\theta),
\end{equation}
where
\begin{equation}
T_1(r) = E\frac{(r^2+a^2)^2}{\Delta} + aL\left(1-\frac{r^2+a^2}{\Delta}\right),\ T_2(\theta) = -Ea^2\sin^2\theta.
\end{equation}
Thus,
\begin{equation}
\begin{aligned}
\Gamma & = \frac{1}{\Lambda_r}\int_0^{\Lambda_r} d\lambda\, T_1 + \frac{1}{\Lambda_\theta}\int_0^{\Lambda_\theta} d\lambda\, T_2\\
& = \frac{2}{\Lambda_r}\int_{r_{min}}^{r_{max}} \frac{dr}{R(r)^{1/2}} T_1(r) + \frac{4}{\Lambda_\theta}\int_{\theta_{min}}^{\pi/2} \frac{d\theta}{\Theta(\theta)^{1/2}} T_2(\theta),
\end{aligned}
\end{equation}
and we have
\begin{equation}
\Omega_\theta = \Upsilon_\theta/\Gamma = 2\pi/(\Lambda_\theta\Gamma),\ \Omega_r = \Upsilon_r/\Gamma = 2\pi/(\Lambda_r\Gamma).
\end{equation}

Similarly,  $\Phi(r,\theta)$ can be written as
\begin{equation}
\Phi(r,\theta) = \Phi_1(r) + \Phi_2(\theta),
\end{equation}
where
\begin{equation}
\Phi_1(r) = aE\left(\frac{r^2+a^2}{\Delta}-1\right) - L\frac{a^2}{\Delta},\ \Phi_2(\theta) = \frac{L}{\sin^2\theta}.
\end{equation}
Thus,
\begin{equation}
\begin{aligned}
\Upsilon_\phi =& \frac{1}{\Lambda_r}\int_0^{\Lambda_r} d\lambda\, \Phi_1 + \frac{1}{\Lambda_\theta}\int_0^{\Lambda_\theta} d\lambda\, \Phi_2\\
=& \frac{2}{\Lambda_r}\int_{r_{min}}^{r_{max}} \frac{dr}{R(r)^{1/2}} \Phi_1(r) + \frac{4}{\Lambda_\theta}\int_{\theta_{min}}^{\pi/2} \frac{d\theta}{\Theta(\theta)^{1/2}} \Phi_2(\theta),
\end{aligned}
\end{equation}
and we have
\begin{equation}
\Omega_\phi = \Upsilon_\phi/\Gamma.
\end{equation}

\bibliographystyle{utphys}
\bibliography{Note}

\end{document}